\documentclass[aps,prl,preprintnumbers,amsmath,amssymb,latexsym,nofootinbib,array,enumerate,letter,twocolumn,superscriptaddress]{revtex4-1}
\usepackage{blindtext}
\usepackage{lineno}
\usepackage{bm,color,xcolor}
\usepackage{slashed} 
\usepackage{graphicx}
\usepackage{subfigure} 
\usepackage{amsmath}
\usepackage{amssymb}
\usepackage{multirow}
\usepackage{float}
\usepackage{comment}
\usepackage{booktabs}
\usepackage{mathrsfs}
\usepackage{fontawesome} \usepackage{mathtools}
\usepackage[normalem]{ulem}
\usepackage[colorlinks=true
,urlcolor=blue
,anchorcolor=blue
,citecolor=blue 
,filecolor=blue
,linkcolor=blue
,menucolor=blue
,linktocpage=true
,pdfproducer=medialab
,pdfa=true
]{hyperref}
\usepackage{lipsum}

\makeatletter
\def\l@subsubsection#1#2{}
\makeatother

\newcommand{\MeV}{\ensuremath{\mathrm{MeV}}}
\newcommand{\GeV}{\ensuremath{\mathrm{GeV}}}

\newcommand{\brr}[1]{\left(#1\right)}
\newcommand{\srr}[1]{\left[#1\right]}

\begin{document}

\title{Electroweak Baryogenesis from Collapsing Domain Walls}

\author{Yang Bai}
\email{yangbai@physics.wisc.edu}
\affiliation{Department of Physics, University of Wisconsin-Madison, Madison, WI 53706, USA} \affiliation{HEP Division, Argonne National Laboratory, 
Argonne, IL 60439, USA}

\author{Kun-Feng Lyu} 
\email{kunfeng.lyu@ou.edu}
\affiliation{Homer L. Dodge Department of Physics and Astronomy, University of Oklahoma, Norman, OK 73019, USA}

\author{Yue Zhao}
\email{zhaoyue@ust.hk}
\affiliation{Department of Physics and Jockey Club Institute for Advanced Study, The Hong Kong University of Science and Technology, Hong Kong S.A.R., P.R.China}

\begin{abstract}
We propose a novel mechanism for electroweak baryogenesis in which collapsing domain walls formed by an axion-like field replace the bubble walls in a strong first-order electroweak phase transition. The axion-like particle coupling to the Higgs mass term allows domain walls to separate regions with distinct electroweak phases, while the electroweak crossover induces a potential-energy bias that triggers their collapse. The directed wall motion, through the axion-like particle coupling to the electroweak topological term, acts as an effective baryon chemical potential and generates an asymmetry via electroweak sphaleron processes. We show that the observed baryon asymmetry can be obtained from either late-time entropy injection or sphaleron suppression in a weakly broken electroweak domain. The wall collapse also produces a stochastic gravitational-wave background with features distinct from standard electroweak-scale first-order-transition spectra.
\end{abstract}

\maketitle

\setcounter{secnumdepth}{3}
\setcounter{tocdepth}{1}

\setcounter{topnumber}{5}
\renewcommand{\topfraction}{1.0}
\renewcommand{\textfraction}{0.0}

{\bf Introduction}---%
The origin of the observed baryon asymmetry of the Universe remains one of the most profound puzzles at the interface of particle physics and cosmology. This asymmetry is quantitatively characterized by the baryon-to-photon ratio inferred from cosmic microwave background (CMB) and Big Bang Nucleosynthesis (BBN) observations, $\eta_B \equiv (n_B - n_{\bar{B}})/n_\gamma \approx 6.1 \times 10^{-10}$~\cite{Planck:2018vyg}. 

Among the various baryogenesis models,
electroweak baryogenesis (EWBG)~\cite{Kuzmin:1985mm,Cohen:1993nk,Carena:1996wj,Riotto:1999yt,Trodden:1998ym,Dine:2003ax,Cline:2006ts,Morrissey:2012db} is particularly appealing, as it operates at the electroweak scale and is testable in current and near-future experiments. However, within the Standard Model (SM), the CP-violating phase in the CKM matrix is too small to generate the observed baryon asymmetry~\cite{Huet:1994jb,Gavela:1993ts,Gavela:1994ds,Gavela:1994dt,Gavela:1994yf}. In addition, the electroweak phase transition (EWPT) is a crossover~\cite{Jansen:1995yg,Kajantie:1995kf,Rummukainen:1996sx,Kajantie:1996mn,Gurtler:1997hr,Csikor:1998eu,Laine:1998vn,Laine:1998qk,Rummukainen:1998as}, rather than the strong first-order phase transition (SFOPT) required to suppress sphaleron washout~\cite{Manton:1983nd,Klinkhamer:1984di}. New physics beyond the SM is therefore required to realize EWBG.
However, stringent electric dipole moment (EDM) bounds~\cite{ACME:2018yjb,Roussy:2022cmp} severely constrain new CP-violating sources, while Higgs precision measurements at the LHC~\cite{Chung:2012vg,CMS:2022dwd,ATLAS:2022vkf} disfavor the sizable potential deformations typically needed to realize an SFOPT. As a result, a broad class of electroweak baryogenesis scenarios is now tightly constrained. These challenges have motivated alternative mechanisms, including domain-wall-assisted baryogenesis~\cite{Daido:2015gqa,Azzola:2024pzq,Mariotti:2024eoh,Azzola:2026cwa,Brandenberger:1994mq,Abel:1995uc} and axion-assisted EWBG~\cite{Im:2021xoy,Bhandari:2025phe,Jeong:2024hhi,Jeong:2018jqe,Jeong:2018ucz}.

In this study, we introduce a novel mechanism in which collapsing domain walls serve as the dynamical objects that naturally replace the bubble walls of an SFOPT and realize baryogenesis through electroweak sphaleron processes. More specifically, we consider a scenario in which EWBG is associated with heavy axion-like particle (ALP) domain walls~\cite{Zeldovich:1974uw,Kibble:1976sj,Vilenkin:1984ib,Sikivie:2006ni,Sikivie:1982qv} formed at a high energy scale. The ALP coupling to the Higgs mass term leads to a situation in which electroweak (EW) symmetry is broken in one domain but preserved in another as the Universe cools to below around the electroweak scale. The domain wall (DW) separates the two distinct domains and maintains their different phases until the DW collapses under a potential-energy bias. During the collapse, the motion of the ALP domain wall, together with the ALP anomalous coupling to the weak gauge fields, can generate a net baryon asymmetry through its effective role as a baryon chemical potential. CP symmetry is conserved in this model, which can therefore be viewed as a novel realization of spontaneous baryogenesis~\cite{Cohen:1988kt,Cohen:1991iu}, with the baryon asymmetry generated around the electroweak phase transition temperature. 

In our model, there is no requirement for an SFOPT, which usually requires a nontrivial arrangement of Lagrangian parameters and sometimes unusually large couplings in the scalar potential. By having domain walls replace bubble walls, one can overcome the traditional difficulties while still realizing a similar situation, with electroweak symmetry broken on one side of the wall and preserved on the other. Because of the interplay between the ALP and the Higgs field, the collapse of the domain walls is triggered by the electroweak crossover transition. In this sense, our model still realizes electroweak baryogenesis, but with distinct early-Universe and phenomenological consequences compared to traditional electroweak baryogenesis with an SFOPT.

{\bf Model}---%
Consider a charged complex scalar field $S$, the simplest potential $V(S) = \lambda_S(|S|^2 - f_a^2/2)^2$ can make the $S$ field obtain a non-vanishing vacuum expectation value (VEV) parametrized as 
\begin{equation}
  S(x) = \dfrac{f_a + \rho(x)}{\sqrt{2}} \exp\brr{i\frac{a(x)}{f_a}}  \ .
\end{equation}
Here $\rho(x)$ field is the radial mode and the angular mode $a(x)$ field as a pseudo-Nambu-Goldstone boson is the ALP.  

Let us first break the shift symmetry of the axion field down to a $Z_2$ symmetry by introducing an explicit symmetry-breaking term of the form $m^2 (S^2 + S^{*2})$. This yields an axion potential given by $\Lambda^4 \cos\left(\frac{2a}{f_a}\right)$, where $\Lambda^4 =  m^2 f_a^2$. 
Consequently, there are two degenerate vacua along the axion field direction, located at $0$ and $\pi$.
In addition, we couple this ALP field to the Higgs sector via the $Z_2$-breaking term $\left(\mu^2 H^\dagger H + \Lambda_2^4\right) \cos\left(\frac{a}{f_a} + \bar{\theta}\right)$, with $\bar{\theta}$ as an additional bare phase that could be related to the $\theta$ angle of a hidden-confining gauge group to generate this operator. Here $\Lambda_2^4$ term at least receives the one-loop contribution proportional to $\mu^2$ from the first term by closing the Higgs field (see Appendix~\ref{app:UV} for discussion of a potential UV origin for this term).
As a result, the Higgs mass differs depending on the axion field value. Note that when $\Lambda_2^4 < \mu^2 v^2/2$ with $v=246$~GeV, electroweak symmetry breaking provides the dominant $Z_2$-breaking source.

Combining both contributions, we obtain the total potential for the Higgs and ALP fields as
\begin{equation}\label{eq:tot_V}
\begin{split}
    V(H,a) =& -\mu_H^2  H^\dagger H + \lambda \brr{H^\dagger H }^2 - \Lambda^4 \cos\brr{\frac{2a}{f_a}}  \\
    &- \brr{\mu^2 H^\dagger H + \Lambda_2^4}\cos\brr{\frac{a}{f_a} + \bar{\theta}} \ .
\end{split}
\end{equation}
In our study, we take $\Lambda \gg \mu, \Lambda_2$. Thus, the ALP potential is primarily determined by the $\Lambda^4$ term. Neglecting the Higgs coupling to the ALP, the $\Lambda_2^2$ term in Eq.~\eqref{eq:tot_V} breaks the $Z_2$ degeneracy. The ALP potential with $\bar{\theta} \subset [0,\pi)$ is lower for $\theta_a \equiv a/f_a$ around 0 and higher for $\theta_a$ around $\pi$. 
In these two vacua, the Higgs quadratic terms are approximately given by $- (\mu_H^2 \pm \mu^2 \cos\bar{\theta})$. For $\mu^2 \cos\bar{\theta} > \mu_H^2$, one can have the electroweak symmetry to be unbroken in the metastable vacuum with $\theta_a \approx \pi$, and the normal electroweak symmetry breaking (EWSB) with $\langle H \rangle = v/\sqrt{2}$ and the Higgs boson mass of $m_h^2 = 2\,(\mu_H^2 + \mu^2 \cos{\bar{\theta}})$ in the global vacuum with $\theta_a \approx 0$.

In the normal global vacuum with $\theta_a \approx 0$ and $\langle H \rangle = v/\sqrt{2}$, a mass mixing term between the Higgs boson and the ALP arises. Upon converting to mass eigenstates, the mixing angle is given by 
\begin{equation}\label{eq:mix——angle}
    \theta_{ah} = \mu^2\,v\,\sin\bar{\theta}/[\sqrt{2} f_a (m_a^2 - m_h^2)] \ ,
\end{equation}
through which the ALP can decay into SM particles.

{\bf Cosmic Evolution}---%
We require the existence of ALP domain walls without axion string at high temperature. We do not specify the detailed formation mechanism but anticipate some mechanism similar to the Kibble one.
In the early universe, one can develop two degenerate vacuum states for the ALP field, $\langle\theta_a \rangle= 0$ and $\langle\theta_a \rangle = \pi$. Different correlation-disconnected patches will randomly select certain vacuums. 
We denote $\langle\theta_a \rangle=0$ as 0-domain and $\langle \theta_a \rangle=\pi$ as $\pi$-domain.
As a result, the ALP domain wall forms. The domain wall networks  quickly enter the scaling regime~\cite{Press:1989yh,Garagounis:2002kt,Oliveira:2004he,Leite:2011sc,Blasi:2025tmn} and there are only $\mathcal{O}(1)$ DWs in each Hubble patch. 
At high temperature and ignoring the small $Z_2$-breaking effects, the two domains stand for two degenerate vacuums. 

One can solve the ALP domain wall profile from the static equation of motion. Assuming the axion field only changes along the $z$ direction, one then obtains
\begin{equation}
    a(z) = 2 f_a \arctan \srr{\exp\brr{ m_a\,z}} \ .
\end{equation}
The ALP wall tension is given by 
\begin{equation}
    \sigma_{\rm w} = (1.3\times 10^{7} \GeV)^3\,\brr{\dfrac{m_a}{10 \,\GeV}} \brr{\dfrac{f_a}{10^{10}\,\GeV}}^2 \ .
\end{equation}
 
\begin{figure}[th!]
    \centering
    \includegraphics[width=0.95\linewidth]{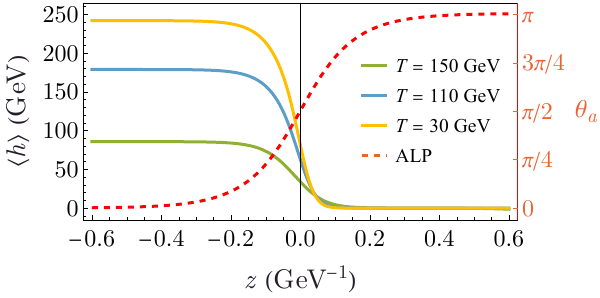}
    \caption{The illustration of the ALP and Higgs field profiles across the domain wall with $m_a = 10\,\GeV$ for different temperatures.}
    \label{fig:wall_profile}
\end{figure}

As the universe cools down below the normal electroweak crossover phase transition temperature $T_c \approx 160\,\mbox{GeV}$~\cite{DOnofrio:2015gop}, the Higgs field in the 0-domain develops a nonzero VEV with $\langle H \rangle = v(T)/\sqrt{2}$, while the $\pi$-domain stays electroweak symmetric. A nonzero Higgs VEV in the 0-domain together with the $\Lambda_2^4$ term in the second line of Eq.~\eqref{eq:tot_V} generates a potential bias $V_{\rm bias} \approx  2\,\Lambda_2^2\cos{\bar{\theta}} + m_h^2 v^2/8$ when $T \ll T_c$. Meanwhile, the nontrivial Higgs profile will develop across the wall. Using the approximation of $v^2(T) = v^2(1-T^2/T_c^2)$, the Higgs wall profile can be solved from the equation of motion under the ALP domain wall background. In Fig.~\ref{fig:wall_profile}, we display the ALP wall as well as the Higgs wall field profiles. The three different colored curves are for the distinct Higgs VEV or at different temperatures. The dashed line refers to the ALP wall profile. These features lead to baryon number production, which will be discussed in the subsequent section.

The domain wall network would start to collapse~\cite{Sikivie:1982qv} if $\sigma_{\rm w}\, H(T) \simeq %\Delta
V_{\rm bias}(T)$ at $T = T_{\rm ann}$. Exploiting the thermal potential, one can show that for $T > T_{\rm ann}$, $\sigma_{\rm w}\,H(T)$ always surpasses the potential bias so that the DW keeps metastable before this annihilation temperature in the parameter space we are interested. 
Since the domain wall network would quickly lose causality once it dominates the universe at the temperature $T_{\rm dom}$, we impose the additional requirement $T_{\rm ann} > T_{\rm dom}$, which gives
$V_{\rm bias} > 3.2\,\sigma_{\rm w}^2/M_{\rm pl}^2$. 
Due to the vacuum pressure, the domain wall moves from the 0-domain towards $\pi$-domain. Since the two domains are no longer degenerate, one can treat the walls including the ALP and the Higgs field profiles as the effective expanding bubbles.
The inter-bubble separation is at the order of the Hubble size for the DW network in the scaling region.
Due to the thermal friction from the particle traversing and bouncing back from the Higgs wall, the wall is expected to reach a terminal velocity and eventually shrink and collide.  
Most of the wall energy is converted to the particles. The Higgs field inside the wall turns to the SM particles and the ALP field inside the wall is fragmented into a large number of ALP particles. A small portion of the energy is transferred to the gravitational waves.

{\bf Electroweak Baryogenesis from Collapsing  Walls}---%
We assume that the ALP has an additional coupling to SM particles through the electroweak topological term~\cite{Harigaya:2022ptp}, which can be realized for the $U(1)_S$ symmetry anomalous under electroweak gauge interaction
\begin{equation}
\frac{\alpha_2}{8\pi} \frac{a}{f_a} W^{\mu\nu} \widetilde{W}_{\mu\nu} \ ,
\end{equation}
$\alpha_2=g^2/4\pi$ with $g$ the $SU(2)_L$ gauge coupling.
As we will show, this coupling enables the evolution of the domain wall to induce baryogenesis.
The mechanism is similar to the traditional local baryogenesis via a dimension‑6 coupling between the Higgs field and the electroweak topological term~\cite{Dine:1990fj}. During the collapse of the wall, its motion generates an effective chemical potential. This leads to a local imbalance between sphalerons and anti‑sphalerons, leading to the rapid generation of a net baryon number.

In SM, both the baryon number $B$ and the lepton number $L$ are anomalous under the electroweak symmetry. The evolution of the baryon number density is therefore given by
\begin{equation}
\frac{dn_B}{dt} =
N_g \left\langle
\frac{\alpha_2}{8\pi} W^{\mu\nu}\widetilde{W}_{\mu\nu}
\right\rangle \,,
\end{equation}
with $N_g = 3$. The thermal average of the Chern–Simons  number density depends both on the driving force $\dot{a}$ and on the energy cost of creating a particle number against a chemical potential~\cite{Moore:1996qs,Berghaus:2020ekh}:
\begin{equation}
\left\langle
\frac{\alpha_2}{8\pi} W^{\mu\nu}\widetilde{W}_{\mu\nu}
\right\rangle = \frac{ \Gamma_{\rm sph}}{2\,T} \left[ \frac{d}{dt}\left(\frac{a}{f_a}\right) - \frac{13}{2} \frac{n_B}{T^2} \right] \,.
\end{equation}
Here $\Gamma_{\rm sph} \simeq 20\, \alpha_2^5\,T^4$~\cite{Moore:1998swa,DOnofrio:2014rug} is the electroweak sphaleron rate per unit volume. The first term inside the parentheses is the axion field velocity, which acts as an effective chemical potential, while the second term describes the washout effect. To a good approximation for a small $n_B$, the washout can be neglected, and we focus on the driving term alone. Furthermore, to avoid washout in the true vacuum, it is required $T_{\rm ann} < 130\,\GeV$~\cite{DOnofrio:2014rug}. 

Right after being swiped by the wall, we can integrate the evolution equation and obtain the local baryon number density
\begin{equation}
n_B \simeq \frac{3\,\Gamma_{\rm sph}}{2\, T\,\gamma} \Delta \theta_a \ ,
\end{equation}
where $\gamma$ is the Lorentz boost factor of the wall. 
According to Fig.~\ref{fig:wall_profile}, we impose a cutoff around $\langle h\rangle \sim \mathcal{O}(100)$ GeV, where the sphaleron rate drops sharply. This corresponds to $\Delta \theta_a \sim 1$. 
It is important to emphasize that this value represents the local baryon number density only in the region swept by the wall. No baryon number is generated in regions that remain in the true vacuum throughout the entire process.

{\bf Baryon Asymmetry}---%
Substituting the explicit values and averaging over the entire space,
the corresponding yield for the baryon number asymmetry is given by
\begin{equation}\label{eq:B_eta_anni}
  Y_B (T_{\rm ann}) \approx \dfrac{n_B}{s} \simeq \dfrac{30\,\alpha_2^5/2}{2\pi^2\,g_{*S} /45} = 1.5\times 10^{-8}  \ ,
\end{equation}
with $\gamma = \mathcal{O}(1)$ and $g_{*S}\approx 100$. Here we assume that the wall reaches a semi-relativistic terminal velocity due to thermal friction from particles scattering off the wall, and that the thermal bath temperature remains approximately at $T_{\rm ann}$.
During the subsequent evolution of the Universe, the baryon-to-entropy ratio $Y_B$
must be further reduced to match the currently observed value, $Y_{B,0} = 0.14\,\eta_{B,0} = 8.4 \times 10^{-11}$. 
This requires a reduction factor of
\begin{equation}
    \mathscr{R} = Y_{B,0}/Y_B(T_{\rm ann}) = 5.6\times 10^{-3} \ .
\end{equation}

Several mechanisms can account for this reduction. Here, we outline two plausible scenarios.
\begin{itemize}
\item \textit{I. Entropy Injection via ALP Decay} \
The ALPs are expected to have a long lifetime before they decay into SM particles. Thus, the Universe may undergo a period of ALP-dominated matter expansion prior to their decay-induced reheating, and the baryon asymmetry can be diluted.

\item \textit{II. Nonzero Higgs VEV in the $\pi$-domain} \
In our initial setup, the false vacuum or the $\pi$-domain is assumed to reside in the electroweak-unbroken phase. Relaxing this assumption by allowing the false vacuum to acquire a small VEV can raise the sphaleron energy barrier and consequently suppress the sphaleron transition rate.
\end{itemize}

In the remainder of this study, we focus on these two approaches~\footnote{One can also consider the possibility of increasing the ALP mass. As the wall becomes thinner, particles spend less time accumulating therefore the baryon asymmetry can be suppressed potentially. While the WKB approximation breaks in the thin wall limit, one should compute the refraction and reflection across the wall.} and investigate their implications in detail. 

\emph{Case I: Entropy Injection via ALP Decay}---%
We assume that the ALP decays only into SM particles, with no additional decay channels. Through its mixing with the Higgs boson, the decay width is given by
$\Gamma_a (m_a) = \Gamma_H(m_a)\,\theta_{ah}^2$
for ALP masses $m_a \sim \mathcal{O}(10\,\GeV)$, which are of primary interest in this study. In the parameter space of interest, $\theta_{ah}^2$ from Eq.~\eqref{eq:mix——angle} is of order $\mathcal{O}(10^{-17})$, far below the current experimental bounds~\cite{Krnjaic:2015mbs,Batell:2022dpx}.
\begin{figure}
    \centering
\includegraphics[width=0.95\linewidth]{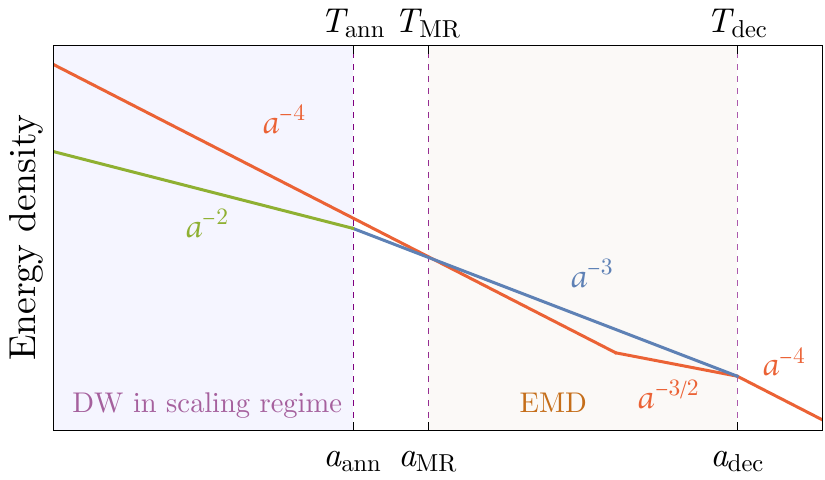}
    \caption{Evolution of the energy densities of different components as functions of the scale factor or radiation temperature during entropy injection. The red, green, and blue curves represent radiation, ALP domain walls, and ALP particles, respectively. Here $T_{\rm ann}$, $T_{\rm MR}$, and $T_{\rm dec}$ are the domain-wall annihilation, early matter-radiation equality, and ALP-decay temperatures, respectively.
    }
    \label{fig:evo_illu}
\end{figure}

Upon collision, the Higgs wall converts into SM radiation, while the ALP wall transforms into ALP particles, which are expected to be semi-relativistic. As a good approximation, the wall energy is almost entirely transferred to the ALP particles. These particles quickly become non-relativistic and begin to dominate the Universe after the SM radiation and ALP matter reach equality at temperature $T = T_{\rm MR}$, corresponding to a Hubble rate $H_{\rm MR}$. Because the couplings between the ALP and SM particles are rather weak, the ALP particles do not enter thermal equilibrium with the SM radiation. Instead, they evolve independently, with a small decay width $\Gamma_a$. As the Universe cools, the ALP particles eventually decay completely when $H(T_{\rm dec}) \simeq \Gamma_a$. BBN constraints require $T_{\rm dec} > 4 \,\MeV$. The explicit evolution history of the Universe is illustrated in Fig.~\ref{fig:evo_illu}. After the Universe enters the early matter-dominated (EMD) stage, the radiation energy density first scales as $a^{-4}$ and subsequently as $a^{-3/2}$ due to entropy injection from ALP decay. The transition occurs when the two contributions become comparable~\cite{Scherrer:1984fd}.

\begin{figure}
    \centering
    \includegraphics[width=0.8\linewidth]{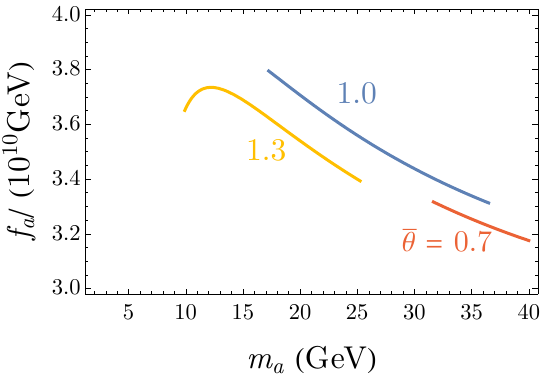} \\ \vspace{3mm}
    \includegraphics[width=0.8\linewidth]{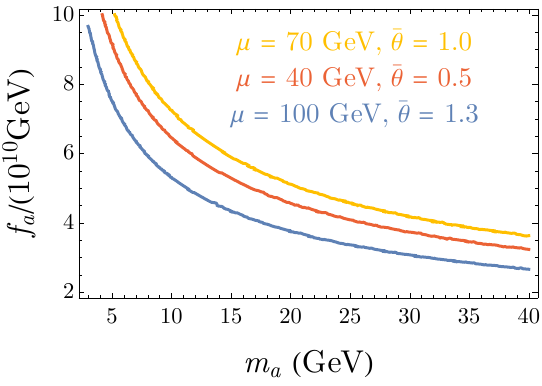}
    \caption{The parameter space producing the observed baryon asymmetry. The upper panel is for case I with entropy injection under $\mu = 150\,\GeV$ and the lower panel is for case II with EWSB in both domains. We choose $\Lambda_2 = 200\,\GeV$ in both cases.
    }
    \label{fig:eta_contour}
\end{figure}

Due to this short EMD epoch, the baryon asymmetry immediately after ALP decay can be expressed as
\begin{align}
Y(T_{\rm dec}) = Y(T_{\rm MR}) \brr{\dfrac{a_{\rm MR}}{a_{\rm dec}}}^{3/4}
\approx
Y(T_{\rm MR}) \brr{\dfrac{H_{\rm MR}}{H_{\rm dec}}}^{-1/2} \ .
\end{align}
From Eq.~\eqref{eq:B_eta_anni}, one finds that achieving the currently observed baryon asymmetry requires
$\Gamma_{a}/H_{\rm MR} = \mathscr{R}^2 = 3 \times 10^{-5}$, with the scaling $\mathscr{R}\sim \mu^2 m_a^{-1} f_a^{-4}$. For illustration, we choose the benchmark values
$\Lambda_2 = 200\,\GeV,\mu = 150\,\GeV$ at the EW scale,
and vary the angle $\bar{\theta}$.
In the upper panel of Fig.~\ref{fig:eta_contour} for case I, the colored curves for three different values of $\bar{\theta}$ represent contours in the $f_a-m_a$ parameter plane that yield the correct baryon asymmetry observed today.

%%%%%%%%%%%%%%%%%%%%%%%%%%%%%%%%%%%%%%%%%
\emph{Case II:  Nonzero Higgs VEV in the $\pi$-domain}---
Another viable possibility for reducing the baryon asymmetry to the observed value is to relax the assumption that the Higgs sector remains completely unbroken in the $\pi$-domain. Instead, we allow the Higgs VEV in this domain to be nonzero but smaller than the SM VEV in the 0-domain.

Let $T_{c,\pi} < T_{c} \simeq 160\ \mathrm{GeV}$ denote the critical temperature for the crossover phase transition in the $\pi$-domain. By choosing the domain wall annihilation temperature $T_{\rm ann}$ to be slightly below $T_{c,\pi}$, the Higgs field can acquire a small VEV. In this broken phase, the suppressed sphaleron rate per unit volume is approximately given by
\begin{equation}
\Gamma_{\rm sph}(T) \simeq 20 \, \alpha_2^5\,T^4 \, e^{-E_{\rm sph}(T)/T}\,,
\end{equation}
where the sphaleron energy is $E_{\rm sph} \simeq 3.34\, m_W(T)/\alpha_2$~\cite{Quiros:1999jp}.
Compared to the symmetric phase, the broken phase introduces an exponential suppression factor due to the sphaleron energy barrier.

In this scenario, achieving the correct baryon asymmetry requires the exponential factor to provide approximately two orders of magnitude of suppression. This yields the condition $E_{\rm sph}(T_{\rm ann})/T_{\rm ann} \simeq 5.2$, which in turn implies that the VEV in the $\pi$-domain at the DW annihilation temperature is $v_\pi(T_{\rm ann}) \approx 0.16\,T_{\rm ann}$.
In the lower panel of Fig.~\ref{fig:eta_contour}, the allowed parameter space is shown by the colored curves for different values of $(\mu, \bar{\theta})$ with $\Lambda_2 = 200\,\GeV$. Note that the relative values of $f_a$ for a fixed $m_a$ depend nontrivially on $\mu$ and $\bar{\theta}$ through $T_{\rm ann}$ and $v_\pi$.

%%%%%%%%%%%%%%%%%%%%%%%%%%%%%%%%%%%%%%%%%
{\bf Gravitational Wave Signal}---
The violent wall collision can also generate gravitational-wave signals. After the collision, sound waves and magnetohydrodynamic (MHD) turbulence could serve as long-lasting sources of GWs.
The peak $\Omega_{\rm GW}$ contributions from sound waves and turbulence are proportional to $v_b^3$ and $v_b$, respectively, with $v_b$ denoting the wall velocity~\cite{Konstandin:2017sat}. Here we neglect these contributions and focus on the ALP wall collapse due to its large tension.
In the radiation-dominated era, the GW energy density is proportional to $G\,\sigma_{\rm w}^2$, where $G$ is Newton's constant. The GW spectrum per unit logarithmic frequency interval has a peak value given by~\cite{Gleiser:1998na,Hiramatsu:2013qaa,Saikawa:2017hiv,Hiramatsu:2012sc,Bai:2023cqj}
\begin{equation}
\left. \frac{d\rho_{\text{GW}}}{d \ln k} \right|_{\text{peak}} = \tilde{\epsilon}_{\text{GW}} G \mathcal{A}^2 \sigma_{\rm w}^2 \ .
\end{equation}
The prefactor $\tilde{\epsilon}_{\text{GW}} = 0.7 \pm 0.4$ is determined from simulations, and $\mathcal{A} \approx 0.8$ for the $\mathbb{Z}_2$ wall. The peak frequency at collision is approximated by the Hubble scale, $f_{\text{peak}}(t_{\text{ann}}) \approx H(t_{\text{ann}})$.
Motivated by numerical simulations, we adopt the spectral function parametrization $\mathcal{S}(x) = (a + b)^c / (bx^{-a/c} + ax^{b/c})^c$ with $a =3$ and $b \approx c \approx 1$. The IR tail scales as $k^3$ for $k < k_{\rm peak}$ due to causality, while the UV part scales as $k^{-1}$ for $k > k_{\rm peak}$.
\begin{figure}[t]
\vspace{0.1cm}
    \centering
    \includegraphics[width=1\linewidth]{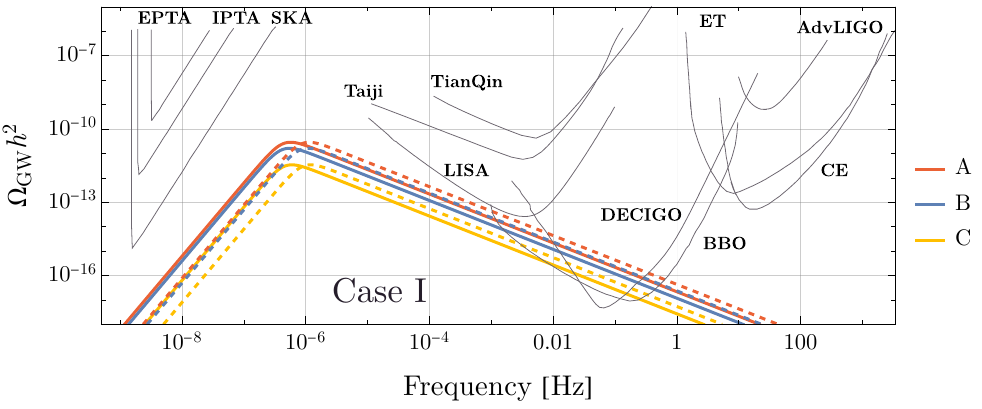} \\ \vspace{3mm}
    \includegraphics[width=1\linewidth]{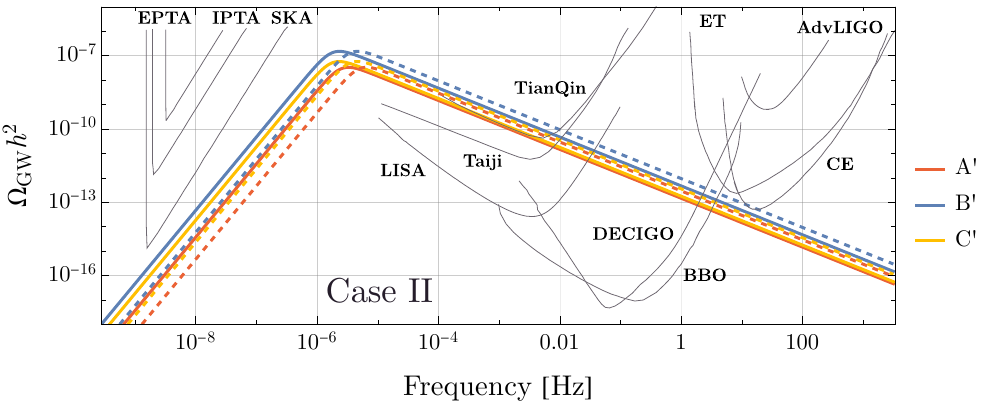}
    \caption{The energy density spectrum of the GW signal for three different benchmark parameter choices. Solid and dashed lines correspond to an average of one and two DWs per Hubble patch, respectively.}
    \label{fig:GW}
\end{figure}
The peak frequency we observe today is given by
\begin{align}
  f_{\rm peak} \approx \,  & 2.6 \times 10^{-6}  {\rm Hz}  \brr{\dfrac{T_{\rm ann}}{100\,\GeV}} 
  \brr{\dfrac{g_*(T_{\rm ann})}{100}}^{1/3}
  \nonumber\\
  & \times \left\{ 
\begin{array}{ll}
    \brr{\dfrac{H_{\rm dec}}{H_{\rm MR}}}^{1/6} \brr{\dfrac{g_{*}(T_{\rm MR})}{g_*(T_{\rm dec})}}^{1/12} & \text{I} 
    \vspace{0.2cm}  \\
    1 & \text{II}
\end{array}
\right.
   \ .
\end{align}
For the entropy injection case I with an EMD stage, $\Omega_{\rm GW}$ is diluted by the extra factor of
$[H(T_{\rm dec})/H(T_{\rm MR})]^{2/3} = \mathscr{R}^{4/3} = 10^{-3}$. 
While for the case II with EWSB in both domains, there is no extra suppression for the GW signal.
Compared to the ordinary EWPT scenario with $\beta/H\sim \mathcal{O}(100)$, this peak frequency is much lower and the GW signal magnitude is much larger since the ``effective" $\beta/H$ takes $\mathcal{O}(1)$ value. 
We plot the observed GW spectrum today in Fig.~\ref{fig:GW} under three distinct benchmark values as shown in Table~\ref{Tab:para} with the expected sensitivity of the future GW observatories. 
In case I, the GW signal can only be detected by BBO~\cite{Crowder:2005nr} and DECIGO~\cite{Sato:2017dkf}. While in case II without dilution, the GW signal is large enough, falling into the reach window of LISA~\cite{LISA:2017pwj}, Taiji~\cite{Ruan:2018tsw}, Tianqin~\cite{TianQin:2015yph,TianQin:2020hid}. 

\begin{table}[t]
\centering
\setlength{\tabcolsep}{6pt}
\renewcommand{\arraystretch}{1.3}
\begin{tabular}{c| c| c| c| c| c| c}
\toprule
 & $A$ & $B$ & $C$ & $A'$ & $B'$ & $C'$ \\ \hline \hline
\midrule
$m_a$ [GeV] 
& 35 & 20 & 10 & 20 & 15 & 10 \\ \hline
$f_a$ [$10^{10}$ GeV] 
& 3.26 & 3.70 & 3.67 & 4.50 & 5.90 & 5.30 \\ \hline
$\mu$ [GeV] 
& 150 & 150 & 150 & 40 & 70 & 100 \\ \hline
$\bar{\theta}$ 
& 0.7 & 1.0 & 1.3 & 0.5 & 1.0 & 1.3 \\ \hline
\bottomrule
\end{tabular}
\caption{Benchmark parameters for the gravitational-wave spectrum with $\Lambda_2 = 200\,\GeV$.}
\label{Tab:para}
\end{table}

{\bf Baryon Inhomogeneity}---%
Baryon number is produced only in the regions swept by the domain walls. Consequently, no baryon asymmetry is generated in regions that remain inside the 0-domain during the collapse. This leads to an inhomogeneous distribution of baryon number on scales comparable to the Hubble radius at $T = T_{\rm ann}$. Such large-scale inhomogeneity cannot be smoothed out by diffusion, given the short mean free paths of protons and neutrons. During BBN, baryons are converted into light elements, and current observations of the deuterium-to-hydrogen (D/H) ratio place tight constraints on the scale of such inhomogeneities~\cite{Bagherian:2025puf, Azatov:2026sdm}.

However, the inhomogeneity may be reduced by fluid advection arising from subsequent sound-wave and turbulence effects. In particular, sound shells are generated around the walls, with properties that depend on the wall velocity. Immediately after collision, these sound shells propagate through the plasma, and the velocity field can be approximated as a superposition of individual shells, with an root-mean-square velocity $\overline{U}_f \sim \mathcal{O}(0.1)$~\cite{Hindmarsh:2013xza,Hindmarsh:2015qta,Hindmarsh:2017gnf,Caprini:2019egz}. Within a Hubble time, the system transitions from the acoustic phase to turbulence, as longitudinal modes convert into rotational ones, forming shocks and eddies. Since the domain-wall separation is $\mathcal{O}(H^{-1})$ in the scaling regime, the fluid motion can persist for several Hubble times.

The baryon-number evolution during this period follows the diffusion-advection equation
\begin{equation}
\frac{\partial n_B}{\partial t} + \nabla \cdot (n_B \boldsymbol{v}_f) - D\,\nabla^2 n_B = 0 \,,
\end{equation}
where $D \sim 6/T$ and $\boldsymbol{v}_f$ is the local fluid velocity. We characterize the typical magnitude of $\boldsymbol{v}_f$ by the RMS value $\overline{U}_f \simeq \sqrt{\langle |\boldsymbol{v}_f|^2 \rangle}$. In the presence of sustained bulk flow, advection dominates over diffusion, leading to an effective transport length $\ell_{\rm adv} \sim \overline{U}_f\, H^{-1}$, rather than the shorter standard diffusion length $\ell_{\rm diff} \sim \sqrt{D\,H^{-1}}$.

As a result, the baryon number can be efficiently redistributed over Hubble scales before the fluid motion dissipates. For two domain walls per Hubble patch and $\overline{U}_f \sim \mathcal{O}(0.5)$, the baryon asymmetry is expected to be sufficiently homogenized to satisfy BBN constraints. Increasing $N_{\rm DW}$ or introducing junction structures can further reduce the required $\overline{U}_f$ and therefore suppress the inhomogeneity scale more efficiently.

{\bf Conclusions}---%
In this Letter, we propose electroweak baryogenesis assisted by ALP domain walls.
Unlike conventional scenarios that rely on an SFOPT from thermal tunneling, ALP domain walls separate the thermal plasma into regions with distinct electroweak phases. The Higgs profile is correlated with the ALP wall, forming an aligned configuration. Through the coupling of the ALP to the electroweak topological term, the directed motion of the wall induces a local chemical potential, generating a net baryon asymmetry as it sweeps through the plasma.

We further investigate two explicit realizations capable of reproducing the observed baryon asymmetry: entropy injection and unequal EWSB in the two domains. As a concrete prediction of this model, we demonstrate that the collapse of the ALP domain walls can produce sizable gravitational-wave signals, potentially observable at future gravitational-wave detectors.

{\bf Acknowledgement}---%
We thank Ting-Kuo Chen for discussion on the sensitivities of gravitational wave experiments.
YB is supported by the U.S. Department of Energy under the contract DE-SC-0017647 and DE-AC02-06CH11357 at Argonne National Laboratory.

{\bf Note Added: }%
While finalizing this paper, Ref.~\cite{Vanvlasselaer:2026fay} appeared, which presents a similar mechanism exploiting ALP domain walls and spontaneous baryogenesis. However, we consider $N_{\rm DW} =2$ rather than $N_{\rm DW} =1$. Our ALP--Higgs coupling respects a UV discrete symmetry rather than being linear, and the loop-induced potential bias is included and can dominate over the Higgs contribution.

\appendix

\section{UV Completion} \label{app:UV}

The UV completion that generates this $\mathbb{Z}_2$-breaking term is given in~\cite{Graham:2015cka,Espinosa:2015eda,Jeong:2018ucz}. We briefly review it here.
Introducing additional vector-like lepton doublets
$L + L^c$ and singlets $N + N^c$, which are charged under the
hidden QCD group confining at $\Lambda_c$, the relevant couplings are
%%%
\begin{equation}
y\, H L N^{c} \;+\; y'\, H^{\dagger} L^{c} N \;+\; m_{L}\, L L^{c} \;+\; m_{N}\, N N^{c}\,.
\label{eq:lagrangian_terms}
\end{equation}
%%%
We can choose $m_N \ll \Lambda_c < m_L$ so that one can integrate out the heavy lepton to obtain the effective mass term for the singlet $N$, namely,
$\left( m_{N} + \frac{y\,y'}{m_{L}}\, |H|^{2} \right) N N^{c}$.
Suppose the effective theta angle in the hidden QCD group is $\bar{\theta}$. Then the condensation of the hidden QCD group leads to
$\Lambda_2^4 \simeq m_N \Lambda_c^3, \quad \mu^2 \simeq y \, y' \Lambda_c^3/m_L$.
It should be noted that closing the Higgs loop of the $|H|^2 \cos(a/f_a+\bar{\theta})$ term generates a quantum correction, which can be estimated as
\begin{equation}
\Delta V_{\rm loop} = \dfrac{y \, y'}{16\pi^2} m_L \log\brr{\frac{\Lambda_c}{m_N}} \Lambda_c^3 \,\cos(a/f_a+\bar{\theta})\, .
\end{equation}
If the bare mass of this singlet approaches zero, $\Lambda_2^4$ can naturally be of the order of $\Delta V_{\rm loop}$.
In the main text, we choose the benchmark value
\begin{equation}
\Lambda_2^4 \simeq \dfrac{\mu^2 m_L^2}{16\pi^2} ~.
\end{equation}

\bibliography{ref}

\end{document}